\documentstyle[epsf,aps,multicol]{revtex}

\begin{document}
\draft
\title {Cavity-induced coherence effects in spontaneous emission from 
pre-Selection of polarization}

\author {Anil K. Patnaik
and G. S. Agarwal\footnote{also at Jawaharlal Nehru Center for Advanced Scientific Research, Bangalore, India}}
\address {Physical Research Laboratory, Navrangpura, Ahmedabad-380 009, India}
\date{October 7, 1998}

\maketitle 
\begin{abstract} 
Spontaneous emission can create coherences in a multilevel atom having
close lying levels, subject to the condition that the atomic dipole matrix
elements are non-orthogonal. This condition is rarely met in atomic
systems.  We report the possibility of bypassing this condition and
thereby creating coherences by letting the atom with orthogonal dipoles to
interact with the vacuum of a pre-selected polarized cavity mode rather
than the free space vacuum. We derive a master equation for the reduced
density operator of a model four level atomic system, and obtain its
analytical solution to describe the interference effects.  We report the
quantum beat structure in the populations.
\end{abstract}

\pacs{PACS No. : 42.50.Ct, 42.50.Md}

\begin{multicols}{2}
\section{Introduction}
	It is well known that the decay of close lying states in atomic
systems can be quite different from that of the decay of an isolated state
\cite{gsabook,qbeat1,qbeat2,lambda,wee,quench1,quench2,emission,lwi,mixing}.  
This is because in the former case the transition amplitudes arising from
each state can interfere with each other. This interference occurs
provided the transition dipole matrix elements ($\vec{d}_{\alpha\beta}$)
satisfy certain conditions\cite{gsabook}.  To be more specific, let us
consider two excited states $|i\rangle$ and $|j\rangle$ decaying to a
common ground state $|g\rangle$. The condition for the interference
between the two decay channels is
\begin{equation}
\vec{d}_{ig} . \vec{d}_{jg}^* \ne 0.
\end{equation}
As a consequence of (1) the populations and coherences get coupled in the 
density matrix equation \cite{gsabook}:
\begin{equation}
\frac{\partial\rho_{ii}}{\partial t} = 
-2\gamma_i \rho_{ii} - (\Gamma\rho_{ji} +\Gamma^*\rho_{ij})~~;
\end{equation}
where $2\gamma_i$ is the decay rate of the level $|i\rangle$ and $\Gamma
\propto \vec{d}_{ig}.\vec{d}_{jg}^*$. This coupling leads to some
remarkable consequences as discussed in various references
\cite{gsabook,qbeat1,qbeat2,lambda,wee,quench1,quench2,emission,lwi,mixing,nuclear}.  
For example, such coupling leads to quantum beats \cite{qbeat1,qbeat2},
phase dependent line shapes \cite{lambda,wee,emission}, quenching of
spontaneous emission \cite{quench1,quench2}, lasing without inversion
\cite{lwi}, and interference in decay of nuclear levels \cite{nuclear}
etc.

	The question arises - what are the systems for which the condition
(1) holds ? Consider for example the $j = 1 \rightarrow j=0$ transition in
an atomic system. Let $|i\rangle$, $|j\rangle$ and $|g\rangle$ in the
above example denote the states $|j=1,m=1\rangle$, $|j=1, m=-1\rangle$ and
$|j=0,m=0\rangle$ respectively. In this case, simple algebra shows that
\begin{equation}
\langle 0,0|\vec{d}|1,1\rangle . \langle 1,-1|\vec{d}|0,0\rangle =
- |d|^2 (\hat{x} + i\hat{y}).(\hat{x} - i\hat{y})^* = 0,
\end{equation}
where $d$ is the reduced dipole matrix element. Thus, the interference
between two decay channels $|1,1\rangle \rightarrow |0,0\rangle$ and
$|1,-1\rangle \rightarrow |0,0\rangle$ will not occur. Xia {\it et. al.}
\cite{quench1} found states in Sodium dimer where the spin-orbit coupling
makes the dipole matrix elements non-orthogonal as the states get mixed.
Several proposals have been made \cite{mixing} to obtain non-orthogonality.
However, it is desirable to examine how the condition (1) can be
{\it bypassed}. The condition (1) arises from the fact that spontaneous
emission occurs in two orthogonal modes of polarization. Therefore if we
{\it pre-select} the polarization mode, then we do not need the condition
(1) for interference to occur. In order to pre-select the polarization, we
consider the problem of spontaneous emission in a mode selective cavity.
It is of course known that the cavity can provide a good way to manipulate
the spontaneous emission from an excited atom \cite{cavity}.

\begin{figure}
\epsfxsize 1.3 in
\centerline{
\epsfbox{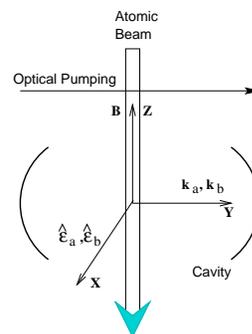}}
\narrowtext{
\caption{
A possible configuration for the pre-selection of polarizations of the
cavity modes that can give rise to new coherences.  The propagation
vectors of the cavity modes $\vec{k}_a, \vec{k}_b$ are along the
Y-direction and cavity polarizations $\hat{\epsilon}_a, \hat{\epsilon}_b$
are along the X-direction, with the quantization axis (Z-direction) fixed
by the direction of the magnetic field $\vec B$.
}}
\end{figure}

\begin{figure}
\epsfxsize 2.3 in
\centerline{
\epsfbox{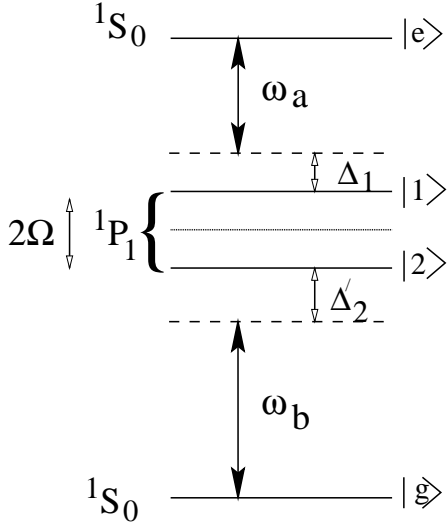}}
\narrowtext{
\caption{
A four-level model scheme (say of $^{40}Ca$) with closely lying
intermediate levels $|1\rangle \equiv |j=1,m=1\rangle$ and
$|2\rangle\equiv |j=1,m=-1\rangle$. Here $\omega_a~(\omega_b)$ is the
frequency of the cavity field coupling $|e\rangle$ to $|1\rangle$ and
$|2\rangle$ ($|1\rangle$ and $|2\rangle$ to the state $|g\rangle$).
$2\Omega$ is the spacing between intermediate levels and the various
detunings are defined by $\Delta_j = \omega_{ej} - \omega_a$,
$\Delta_j^\prime = \omega_{jg} - \omega_b$.
}}
\end{figure}

	In this paper, we demonstrate the possibility of restoring quantum
interference effects in spontaneous emission of an excited atom inside a
cavity with its modes selected suitably, and thus avoid the condition (1).
A possible configuration is shown in Fig.1.  In Section II, we describe
the model atom which consists of two near-degenerate intermediate levels
and orthogonal dipoles. The atom interacts with the cavity modes which are
selected a priori. We consider the bad cavity limit and derive a master
equation which shows evolution of quantum coherence between the degenerate
or near degenerate levels.  We obtain quantum beats in the populations of
the intermediate states as well as the ground state. We present the
solution of the master equation in Section III. We observe a decrease in
the ground state population for some range of parameters. We compare these
results obtained with and without interference terms. In Section IV, we
show that, suitable selection of cavity polarization plays vital role in
determining the occurrence of interference.  Finally in Section V, we make
some concluding remarks.

\section{Dynamics of a Four Level System in a Cavity}
	We consider a two-mode cavity containing a four-level atomic
scheme with say, two near-degenerate Zeeman split magnetic sub-levels
$|1\rangle \equiv |j=1,m=1 \rangle$ and $|2\rangle \equiv|j=1,m=-1\rangle$
as its intermediate states (shown in Fig.2). The ``$a$-mode" (``$b$-mode")  
couples $|e\rangle \leftrightarrow |m\rangle$ ($|m\rangle \leftrightarrow
|g\rangle$) transitions (for $m=\pm 1$). The scheme could be
$^{40}Ca$-cascade, as shown by the symbols in the left hand side of the
figure.  The total Hamiltonian for the atomic system, and the cavity
fields is
\begin{equation}
H = H_A + H_F + H_{AF},
\end{equation}
where,
\begin{eqnarray}
H_A &=& \hbar (\omega_{eg} A_{ee} +\omega_{1g} A_{11} +\omega_{2g} A_{22}) ,
\nonumber\\ 
H_F &=& \hbar (\omega_a a^\dag a +\omega_b b^\dag b),
\nonumber\\ 
H_{AF} &=& - \vec{d}.\vec{E}_{cav} \\ \nonumber
&=& -i\hbar\sum_{j=1}^2 \left( G_{je} a^\dag A_{je} + G_{gj} b^\dag 
A_{gj} \right) + h.c.~.
\end{eqnarray}
Here state $|g\rangle$ is assumed to be the ground state;  $\hbar
\omega_{jg} = ({\cal E}_j - {\cal E}_g)$ (for $j \equiv e, 1, 2$) defines
the energy of the atomic state $|j\rangle$ with respect to $|g\rangle$ and
$A_{ij}=|i\rangle\langle j|$ are the atomic operators that denote
populations (coherences) for $i = j$ ($i \ne j$).  Further, $a, b$
($a^\dag, b^\dag$) are annihilation (creation) operators for the cavity
field modes with frequencies $\omega_a$ and $\omega_b$ respectively.
$E_{cav}$ is the quantized two mode cavity field. The atom-cavity mode
coupling constants are given by
\begin{equation}
G_{je} = 
\left( \frac{2\pi\hbar\omega_a}{V}\right)^{1/2} \frac{\vec{d}_{je}.
\hat{\epsilon}_a}{\hbar}, 
G_{gj} = 
\left( \frac{2\pi\hbar\omega_b}{V}\right)^{1/2} \frac{\vec{d}_{gj}.
\hat{\epsilon}_b}{\hbar}, 
\end{equation}
with $V$ being the cavity volume and $\hat{\epsilon}_a$ and $\hat{
\epsilon}_b$ being the polarizations of the cavity modes.  We work
in the interaction picture. The Hamiltonian in the interaction picture 
is given by
\begin{eqnarray}
H_I (t) &=& e^{\frac{i}{\hbar} (H_A + H_F)t} H_{AF} e^{-\frac{i}{\hbar} 
(H_A + H_F)t}
\nonumber \\
&=& -i\hbar\sum_{j=1}^2 \left( G_{je} a^\dag A_{je} e^{-i\Delta_j t} 
+ G_{gj} b^\dag A_{gj} e^{-i\Delta_j^\prime t} \right) 
\nonumber\\
& &+ h.c. , ~~j=1,2;
\end{eqnarray}
where, $\Delta_j = \omega_{ej} - \omega_a$ and $\Delta_j^\prime =
\omega_{jg} - \omega_b$ are the detunings. The above Hamiltonian describes
the reversible interactions between the atom and the cavity field. However
we should also take into account the irreversible processes due to
coupling of the cavity field with the outside world via cavity mirrors. We
denote the leakage rates of the photons in the cavity modes by $\kappa_a$
and $\kappa_b$. At optical frequencies we can neglect the thermal photons.
We further work in the bad cavity limit. The density matrix equation in
the the interaction picture for the combined atom-field system contains
two parts:  (a) coherent evolution described by the Liouvillian $\Lambda$,
and (b) the field relaxation part described by $\Lambda_F$ \cite{master}
\begin{equation}
\frac{\partial\rho}{\partial t} = (\Lambda + \Lambda_F) \rho,
\end{equation}
where,
\begin{eqnarray}
\Lambda\rho &=& -\frac{i}{\hbar} \left[ H_I(t), \rho \right], 
\\ \nonumber 
\Lambda_F\rho &=&
-\kappa_a (a^\dag a\rho - 2 a\rho a^\dag +\rho a^\dag a)
\nonumber \\
& &-\kappa_b (b^\dag b\rho - 2 b\rho b^\dag +\rho b^\dag b).
\nonumber
\end{eqnarray}

To get useful information about the evolution of the atomic system, we
derive the Master equation for the reduced atomic operator by 
approximately eliminating the cavity field using the standard projection
operator techniques \cite{gsabook,master}. In the following, we outline
some of the important steps. We rewrite Eq.(8) as
\begin{equation}
\frac{\partial{\tilde \rho}}{\partial t} = {\tilde \Lambda} (t) 
{\tilde\rho}(t),
\end{equation}
by transforming to a new frame with the transformations,
\begin{equation}
{\tilde \rho} \equiv e^{-\Lambda_F t} \rho,
~~{\tilde \Lambda} \equiv e^{-\Lambda_F t} \Lambda e^{\Lambda_F t}.
\end{equation}
We define the projection operator to be ${\cal P}...\equiv \rho_F (0) Tr_F
...$ and write Eq.(10) as,
\begin{equation}
\frac{\partial{\tilde \rho}}{\partial t} = 
{\tilde \Lambda} {\cal P}{\tilde \rho} 
+ {\tilde \Lambda} (1-{\cal P}){\tilde \rho}.
\end{equation}
The assumptions that we make are (a) at $t=0$, $\rho(0)$ can be factorised
into a product of atom and field density operators, (b) the photons
emitted can not react back with the atom i.e., we use the Born
approximation and (c) the Markoff approximation $G^2\kappa^{-1} \ll
\kappa$ ($G$ refers to vacuum Rabi frequencies) which ensures that the
system has a short memory. Using (10) and the above approximations and
tracing over the field states Eq.(12) reduces to,
\begin{eqnarray}
\frac{\partial{\rho_a}}{\partial t} &=&-\frac{1}{\hbar^2}
\\ \nonumber
&\times&\lim_{t\rightarrow\infty} 
\int_0^t d\tau Tr_F \left[ H_I(t), e^{\Lambda_F\tau}
[H_I(t-\tau), \rho_F(0){\rho_a}] \right].
\end{eqnarray}
For convenience, $\tilde{\rho_a}$ is replaced by $\rho_a$ in (13) and in 
subsequent calculations. 
	
	The trace over the field operators inside the integral is
calculated using the following relations. For arbitrary field operators
$A$ and $B$, simple trace algebra and the definition of adjoints give
\begin{eqnarray}
Tr_F \left[ A e^{\Lambda_F\tau} B\rho_F(0) \right]
= \langle A(\tau)B \rangle, 
\nonumber \\
Tr_F \left[ A e^{\Lambda_F\tau} \rho_F(0) B\right]
= \langle B A(\tau) \rangle.
\end{eqnarray}
Further, the time correlations for the cavity fields in the absence of the
interaction with the atom are known to be
\begin{eqnarray}
\langle a a^\dag (\tau) \rangle = \langle a(\tau) a^\dag  \rangle =
e^{-\kappa_a\tau},
\nonumber \\
\langle b b^\dag (\tau) \rangle = \langle b(\tau) b^\dag  \rangle =
e^{-\kappa_b\tau},
\end{eqnarray}
with all other second order correlation functions being zero.
 
	Substituting the complete Hamiltonian from Eq.(7) in (13) and
using the relations (14), the trace inside the integral is expressed in
terms of field correlations. Further using (15) and evaluating the
integral in Eq.(13), we obtain the master equation for the atomic density
operator
\end{multicols}
\rule{\hsize}{.1mm}\rule{-.1mm}{.1mm}\rule{.1mm}{2mm}
\begin{eqnarray}
\label{master}
\frac{\partial{\rho}_a}{\partial t} = &-&i(\delta_1 +\delta_2)[A_{ee},
{\rho}_a]
-i [(\delta_1^\prime A_{11} + \delta_2^{\prime} A_{22}), 
\rho_a ] 
\nonumber \\ 
&-& \left\{ \Gamma_1 (A_{ee} {\rho}_a - 2 A_{11} {\rho}_{ee} + {\rho}_a
A_{ee} ) 
+\Gamma_1^{\prime} (A_{11} {\rho}_a - 2 A_{gg} 
{\rho}_{11} + {\rho}_a A_{11} ) + 1\rightarrow 2\right\}
\nonumber\\ 
&+& \left\{ 2 G_{1e} G_{2e}^* \frac{\kappa_a +i\Omega}
{(\kappa_a +i \Delta_2)(\kappa_a - i\Delta_1)} A_{12}{\rho}_{ee} 
e^{2i\Omega t} + h.c. \right\} 
+\left\{ 2 G_{g1} G_{g2}^* \frac{\kappa_b -i\Omega}{(\kappa_b +i\Delta_2^\prime)
(\kappa_b -i\Delta_1^\prime)}A_{gg} {\rho}_{12} e^{-2i\Omega t} +h.c.
\right\} 
\nonumber \\ 
&-& \left\{ G_{g1}^* G_{g2} e^{2i\Omega t}\left( 
\frac{1}{\kappa_b - i\Delta_2^\prime} A_{12} {\rho}_a + 
\frac{1}{\kappa_b + i\Delta_1^\prime} {\rho}_a A_{12} 
\right) + h.c. \right\}
\end{eqnarray}
\widetext
where,
\begin{eqnarray}
\Gamma_j &=& |G_{je}|^2 \kappa_a/(\kappa_a^2 + \Delta_j^2),~~
\Gamma_j^\prime = |G_{gj}|^2 \kappa_b/(\kappa_b^2+ {\Delta_j^\prime}^2),
~\Omega = (\omega_{1g} - \omega_{2g})/2,
\nonumber \\
\delta_j &=&  |G_{je}|^2 \Delta_j/(\kappa_a^2 + \Delta_j^2),~~
\delta_j^\prime = |G_{gj}|^2\Delta_j^\prime/(\kappa_b^2+ {\Delta_j^
\prime}^2), ~~
j=1,2~.
\end{eqnarray}
Here $\Gamma$ and $\Gamma^\prime$'s represent various decay constants from
different levels and $\delta$ and $\delta^\prime$'s are the frequency
shifts of atomic levels resulting from interaction with the vacuum field
in a detuned cavity.  This is the key equation of this paper and will be
used in the subsequent analysis to study the coherence effects induced by
the cavity.

	To understand the meaning of various terms in the Master equation 
(\ref{master}) we write the equations explicitly for the density matrix 
elements: 
\begin{eqnarray}
\label{den-eq}
\frac{\partial\rho_{ee}}{\partial t} = &-&2(\Gamma_1 + \Gamma_2)\rho_{ee},
\nonumber \\
\frac{\partial \rho_{11}}{\partial t} =&-&2\Gamma_1^\prime \rho_{11}
+ 2\Gamma_1 \rho_{ee} 
- \eta ~\frac{G_{g1}^* G_{g2}}{\kappa_b - i\Delta_2^\prime}\rho_{21} 
e^{2i\Omega t}
- \eta ~\frac{G_{g1} G_{g2}^*}{\kappa_b + i\Delta_2^\prime}\rho_{12}
e^{-2i\Omega t},
\nonumber \\
\frac{\partial \rho_{12}}{\partial t} = 
&-&\left( \Gamma_1^\prime +\Gamma_2^\prime +i(\delta_1^\prime -\delta_2^\prime)
\right) \rho_{12} 
+ 2 \eta ~G_{1e}G_{2e}^* \frac{\kappa_a+i\Omega}{(\kappa_a+i\Delta_2)(\kappa_a-
i\Delta_1)}\rho_{ee} e^{2i\Omega t}
\nonumber \\ 
&-&\eta ~G_{g1}^* G_{g2}\left( \frac{\rho_{22}}{\kappa_b-i\Delta_2^\prime} 
+ \frac{\rho_{11}}{\kappa_b +i\Delta_1^\prime} \right) e^{2i\Omega t},
\\ 
\frac{\partial \rho_{gg}}{\partial t}
=&2&\Gamma_1^\prime\rho_{11}  + 2\Gamma_2^\prime \rho_{22}
+ 2\eta ~G_{g1}G_{g2}^* \frac{\kappa_b-i\Omega}{(\kappa_b+i\Delta_2^\prime)
(\kappa_b- i\Delta_1^\prime)}\rho_{12} e^{-2i\Omega t}
+ 2\eta ~G_{g1}^*G_{g2}\frac{\kappa_b+i\Omega}{(\kappa_b-i\Delta_2^\prime)
(\kappa_b+ i\Delta_1^\prime)}\rho_{21} e^{2 i\Omega t}. 
\nonumber 
\end{eqnarray}
\widetext
\begin{multicols}{2}
Equation for $\dot{\rho}_{22}$ is the same as for $\dot{\rho}_{11}$ with
$1\leftrightarrow 2$ and $\Omega \rightarrow -\Omega$. Note the presence
of oscillating components in (\ref{den-eq}). If $\Omega$ is large compared
to damping constants $\Gamma$'s or detunings $\delta$'s, then these
exponentials average out (shown explicitly in the discussion following
Eq.(22)) leading to
\begin{eqnarray}
\frac{\partial\rho_{ee}}{\partial t} = &-&2(\Gamma_1 + \Gamma_2)\rho_{ee},
\nonumber \\
\frac{\partial \rho_{11}}{\partial t} =&-&2\Gamma_1^\prime \rho_{11}
+ 2\Gamma_1 \rho_{ee}, 
\\
\frac{\partial \rho_{12}}{\partial t} = 
&-&\left( \Gamma_1^\prime +\Gamma_2^\prime 
+i(\delta_1^\prime -\delta_2^\prime)\right) \rho_{12}, 
\nonumber \\
\frac{\partial \rho_{gg}}{\partial t}
=&2&\Gamma_1^\prime\rho_{11}  + 2\Gamma_2^\prime \rho_{22}.
\nonumber 
\end{eqnarray}
These equations can be compared with the equations for emission in free
space.  Under the initial condition that the atom is in the state
$|e\rangle$, equations (19) admit simple solutions:
\begin{eqnarray}
\rho_{ee}(t) &=& \exp[{-2(\Gamma_1 + \Gamma_2)t}],
\nonumber \\
\rho_{11} (t) &=& \frac{\Gamma_1}{\Gamma_1 + \Gamma_2 -\Gamma_1^\prime}
\left(\exp[{-2\Gamma_1^\prime t}] - \exp[{-2(\Gamma_1^\prime
+\Gamma_2^\prime) t}] \right),
\nonumber \\
\rho_{gg}(t) &=& 1 - \rho_{ee} - \rho_{11}(t) - \rho_{22}(t),
\end{eqnarray}
and $\rho_{22}(t)$ is same as $\rho_{11}(t)$ with $1\leftrightarrow 2$.

	For $\Omega$ comparable to $\Gamma$'s and $\Delta$'s, the
exponential terms are important. The dynamical equations involve coupling
of populations to coherences and vice-versa. Such couplings give rise to
new coherence effects. Accordingly, let us introduce an interference
parameter $\eta$ in Eq.(\ref{den-eq}), so that $\eta =1(=0)$ would refer
to the presence (absence) of coherence effects.

\section{Quantum Coherences and Solution of The Master Equation}

\noindent
{\it (a) Cavity Induced Intermediate State Coherence:}
	
	It is clear from Eq.(\ref{den-eq}) that, for $\eta = 0$, the
coherence between the intermediate levels is never established; i.e.
$\rho_{ij}=0$ for all times. When $\eta$ is unity, there is a two-fold
possibility for the coherence to evolve - (a) the second term in the
equation for $\dot{\rho}_{12}$ causes evolution of coherence due to
coupling of the states $|1\rangle$ and $|2\rangle$ to the excited state by
the cavity vacuum field ``$a$"  and (b) the third term that arises from
the coupling of $|1\rangle$ and $|2\rangle$ to the state $|g\rangle$ by
the cavity vacuum field ``$b$".  The resulting evolution of coherence is
shown in Fig.3.  For degenerate intermediate levels $|j\rangle$ $(j=1,2)$,
and $\omega_a$ ($\omega_b$) in resonance with $|e\rangle\rightarrow
|j\rangle$ ($|j\rangle\rightarrow |g\rangle$) transition, no such
oscillation is seen - though coherence evolves.

\begin{figure}
\epsfxsize 3.8 in
\centerline{
\epsfbox{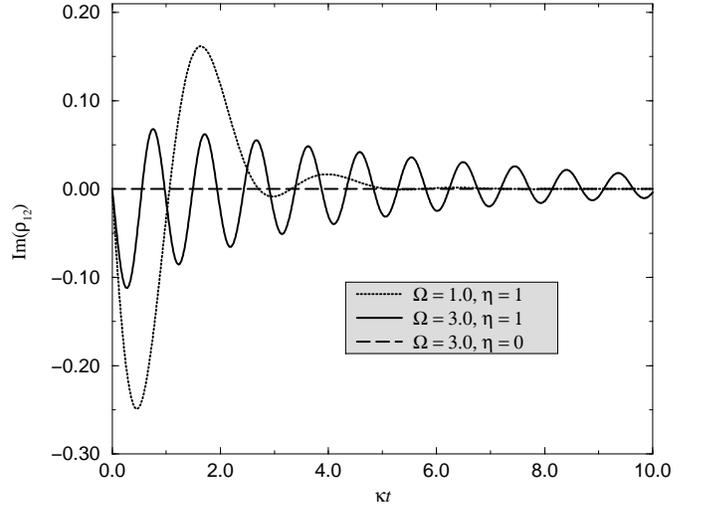}}
\narrowtext{
\caption{
The time evolution of coherence between the intermediate states is
plotted.  All frequencies are scaled with $\kappa_a = \kappa_b = \kappa$.
We choose $G_{je} \equiv G_{gj}=\kappa$, $\Delta_{1}^\prime
=-\Delta_{2}^\prime =\Omega = -\Delta_{1}=\Delta_{2}$. For $\eta = 0$, no
coherence is produced, and for $\eta = 1$, as $\Omega$ increases, the
frequency of oscillation increases but the amplitude of coherence
decreases.
}}
\end{figure}

\noindent
{\it (b) Cavity Induced Quantum Beats in Atomic Populations:}

	For $\eta = 1$, the populations in Eq.(\ref{den-eq}) can be
obtained analytically. For simplicity, assume that $\Gamma_i \equiv
\Gamma_i^\prime \equiv \Gamma$, $G_{ie} \equiv G_{gi} \equiv G$, $\kappa_a
\equiv \kappa_b = \kappa$ and the cavity field $\omega_a(\omega_b)$ is
tuned to the center of the two intermediate states and the excited
(ground) state. Then, the solution of Eq.(\ref{den-eq}) is found to be
\begin{eqnarray}
\label{sol}
\rho_{ii} (&t&) = -\left( 1+\frac{2|\alpha|^2}{\Gamma^2 + f^2}\right) 
e^{-4\Gamma t}
+ \left( 1+\frac{|\alpha|^2}{f^2}\right) e^{-2\Gamma t}
\nonumber \\
&-&\frac{2|\alpha|^2}{\Gamma^2 +f^2} e^{-2\Gamma t} 
\left[ \left(\frac{\Gamma^2}{f^2}- 1\right)\cos (2f t) 
+ \frac{2\Gamma}{f}\sin (2f t) \right], 
\nonumber \\
\rho_{gg}(&t&) = 1 - \rho_{ee}(t) - 2\rho_{ii}(t), i=1,2.
\end{eqnarray}
Here, the parameter $\alpha = GG^*/(\kappa+i\Omega)$ corresponds to the
cross terms in Eq.(\ref{den-eq}). It can therefore be seen that for
$|\alpha| = 0$, Eq.(21) reduces to Eq.(20).  The argument of the
trigonometric functions in Eq.(\ref{sol}) gives the beat frequency
\begin{equation}
2f = 2\left[ (\delta^\prime +\Omega)^2 - |\alpha |^2 \right]^{1/2}. 
\end{equation}
The condition for the beats to occur is $(\delta^\prime +\Omega)^2 > 
|\alpha |^2$. For various values of $\Omega$, we show the time
dependence of $\rho_{ii}$ and $\rho_{gg}$ in Fig.4 assuming $|G|$ to be of
the order of $\kappa$. If the intermediate levels are degenerate ($\Omega
= 0$), then $f$ is purely imaginary and therefore the trigonometric
functions in Eq.(\ref{sol}) change to hyperbolic functions - ceasing the
oscillations in the populations. Again, for $\Omega \ll \kappa$, $f$ is
imaginary and hence there is no beating. However, for $\Omega \sim
\kappa$, the beating in population is prominently seen. An increase in
$\Omega$ leads to increase in the beat frequency. For $\Omega$ very large
compared to $\kappa$, the beat frequency $2f \gg \kappa$ - leading to fast
oscillations, the average of which leads to Eq.(20).
\begin{figure}
\vspace*{0.8 cm}
\epsfxsize 3.3 in
\centerline{
\epsfbox{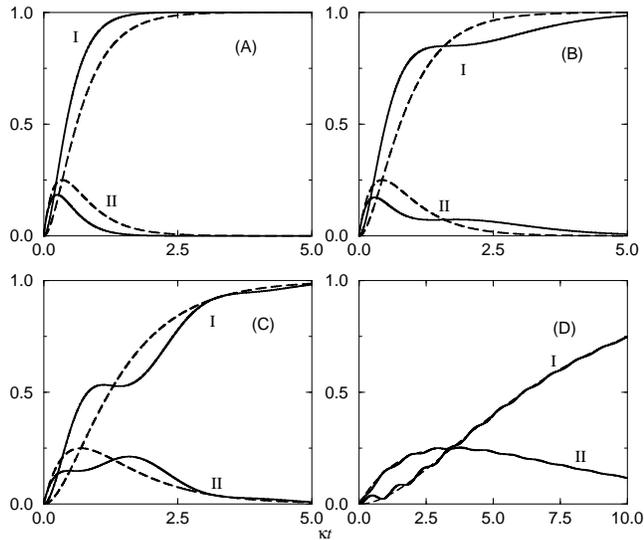}}
\narrowtext{
\caption{
The time dependence of the populations in the ground state $\rho_{gg}$
(represented by I) and the intermediate states $\rho_{11} (=\rho_{22})$
(represented by II). The dashed lines represent $\eta = 0$ where we see no
oscillation. The solid lines represent $\eta = 1$.  The plots for various
values of $\Omega$: (A) $\Omega = 0$ - no beat structure is seen, (B)
$\Omega = 0.5\kappa$, (C) $\Omega = 1.0\kappa$ and (D) $\Omega =
3.0\kappa$ - where the population in the ground state decreases during $t
\sim \kappa^{-1}$.
}}
\end{figure}
 
	Further we note that for $\Omega(\sim 3\kappa)$, the ground state
population decreases for a small time interval implying a population
transfer to the intermediate levels. It should be borne in mind that, we
work in the low-Q cavity limit where cavity vacuum is not strong enough to
cause the vacuum field Rabi oscillation \cite{highQ}. To interpret the
decrease in population, we go back to Eq.(\ref{master}).  The 4th line of
Eq.(\ref{master}) suggests that the ground state population couples the
intermediate state coherences via $G_{g1} G_{g2}^*$ (and $G_{g1}^*
G_{g2}$); e.g., an emission followed by absorption of the same photon on a
different transition. The corresponding transitions would correspond to
$|1\rangle \rightarrow |g\rangle \rightarrow |2\rangle$ (and $|2\rangle
\rightarrow |g\rangle \rightarrow |1\rangle$). The various transitions of
$G_{ij}G_{il}^*$ type and various interference paths are illustrated in
Fig.5. In particular from Fig.5(B), one understands the decrease in the
ground state population.

\begin{figure}
\epsfxsize 3.3 in
\centerline{
\epsfbox{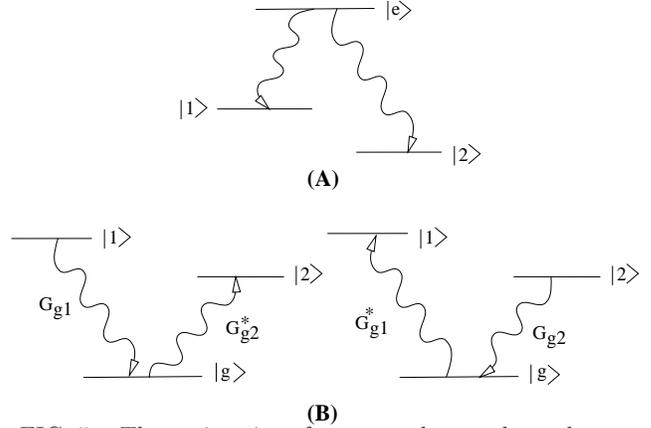}}
\narrowtext{
\caption{
The various interference paths are shown by considering upper and lower
transitions. (A) The upper $\Lambda$ like part: both transitions share a
single reservoir of cavity vacuum - contributing to the coherence between
the states $|1\rangle$ and $|2\rangle$. (B) The lower $V$ like part: to
the lowest order interaction, photons emitted by $|1\rangle
\leftrightarrow |g\rangle$ transition can be absorbed by $|g\rangle
\leftrightarrow |2\rangle$ transition and vice versa - explaining the
decrease in population of state $|g\rangle$.
}}
\end{figure}

\section{Origin of Cavity Induced Coherences}

	We now examine the question - what leads to such coherences which
otherwise do not occur. It is clear from Eq.(\ref{den-eq}) that, the
coherence terms are related to matrix elements like
\begin{equation}
\label{product}
G_{g1} G_{g2}^* = 
\left( \frac{2\pi\omega_b}{\hbar V}\right) 
(\vec{d}_{g1} . \hat{\epsilon}_b) (\vec{d}_{g2}^* . \hat{\epsilon}_b^*). 
\end{equation}
For the choosen geometry of Fig.1, Eq.(\ref{product}) reduces to
\begin{equation}
\label{select}
G_{g1} G_{g2}^* = 
\left( \frac{2\pi\omega_b}{\hbar V}\right) 
(\vec{d}_{g1})_x 
(\vec{d}_{g2}^*)_x. 
\end{equation}
The later is non-vanishing; as for $\sigma_\pm$ transitions, $\vec{d}_{g1}
\equiv -|d|(\hat{x} + i\hat{y})$, $\vec{d}_{g2} = |d| (\hat{x} -
i\hat{y})$. Note further that if polarization can not be pre-selected,
then we have to sum Eq.(\ref{product}) over the two possible polarization
modes leading to 
\begin{eqnarray} \sum_{pol} G_{g1} G_{g2}^* &=& 
\left(\frac{2\pi\omega_b}{\hbar V}\right) 
\sum_{pol} (\vec{d}_{g1} .\hat{\epsilon}_b)  
(\vec{d}_{g2}^* . \hat{\epsilon}_b^*) 
\nonumber \\ 
&=&\left( \frac{2\pi\omega_b}{\hbar V}\right) (\vec{d}_{g1} .\vec{d}_{g2}^*).
\end{eqnarray} 
	Under these conditions the coherence term can survive only
if the dipole matrix elements are non-orthogonal. It is thus clear that,
in order to see the interferences or beats at $2\Omega$, one has to make a
pre-selection of polarization so that coherence between $|1\rangle$ and
$|2\rangle$ can be produced by spontaneous emission. Note that this is
different from the usual quantum beat spectroscopy \cite{spectro,cav-beat} where
coherence is produced by excitation with an {\it external field of
appropriate band width}.

\section{Conclusions}
	In conclusion, we have shown: (a) how the pre-selection of
polarization leads to certain types of interference effects which
otherwise are missing unless the dipole matrix elements are
non-orthogonal; (b) how the pre-selection of polarization can be achieved
in a cavity. We demonstrate this in the context of a four level atomic
system in a bimodal cavity in the limit of a bad cavity. We hope to
consider the effect of the cavity quality on intermediate states
coherences elsewhere.

\end{multicols}

\begin{thebibliography}{99}

\bibitem{gsabook} G.S. Agarwal, ``{\it Quantum Statistical Theories
of Spontaneous Emission and their relation to other approaches}", Springer
Tracts in Modern Physics: Quantum Optics (Springer-Verlag, 1974), Sec.15.

\bibitem{qbeat1} 
D.A. Cardimona, M.G. Raymer, and C.R. Stroud Jr., J. Phys. {B\bf 15}, 
55 (1982); 
A. Imamo\u{g}lu, Phys. Rev. {A\bf 40}, 2835 (1989).

\bibitem{qbeat2} 
G.C. Hegerfeldt, and M.B. Plenio, Phys. Rev. {A\bf 46}, 373 (1992); 
{\it ibid}, {\bf 47}, 2186 (1993); 
Quantum Optics {\bf 6}, 15 (1994);
T.P. Altenm\"{u}ller, Z. Physik {D\bf 34}, 157 (1995).

\bibitem{lambda}
For studies of such coherence effects in the context of $\Lambda$ system see 
J. Javanainen, Europhys. Lett. {\bf 17}, 407 (1992);
S. Menon, and G.S. Agarwal, Phys. Rev. {A\bf 57}, 4014 (1998).

\bibitem{wee}
For studies on $V$-systems, see [1]; 
P. Zhou, and S. Swain, Phys. Rev. Lett. {\bf 77}, 3995 (1996); 
{\it ibid}, {\bf 78}, 832 (1997);
Phys. Rev. {A\bf 56}, 3011 (1997); 
E. Paspalakis, S.Q. Gong, and P.L. Knight, Optics Commn. {\bf 152}, 293 (1998).

\bibitem{quench1} 
H.R. Xia, C.Y. Ye, and S.Y. Zhu, Phys. Rev. Lett. {\bf 77}, 1032 (1996);
G.S. Agarwal, Phys. Rev. {A\bf 55}, 2457 (1997).

\bibitem{quench2} 
S.Y. Zhu, R.C.F. Chan, and C.P. Lee, Phys. Rev. {A\bf 52}, 710 (1995);
S.Y. Zhu, and M.O. Scully, Phys. Rev. Lett. {\bf 76}, 388 (1996);
H. Lee, P. Polynkin, M.O. Scully, and S.Y. Zhu, Phys. Rev. {A\bf 55}, 
4454 (1997).

\bibitem{emission}
M.A.G. Martinez, P.R. Herczfeld, C. Samuels, L.M. Narducci, and C.H. Keitel, 
Phys. Rev. {A\bf 55}, 4483 (1997);
E. Paspalakis, and P.L. Knight, Phys. Rev. Lett. {\bf 81}, 293, (1998). 

\bibitem{lwi}
A. Imamo\u{g}lu, and S.E. Harris, Opt. Lett. {\bf 14}, 1344 (1989);
S.E. Harris, Phys. Rev. Lett. {\bf 62}, 1033 (1989). 

\bibitem{mixing}
We also note that several other methods have been suggested in
the literature to overcome the problem of orthogonal dipole matrix elements.
These include the application of d.c., r.f., or even optical fields depending
on the situation at hand. Here the non-orthogonality is obtained from
mixing of energy levels. The details can be found in
H. Schmidt, and A. Imamo\u{g}lu, Opt. Commn. {\bf 131}, 333 (1996);
A. K. Patnaik, and G. S. Agarwal, J. Mod. Opt. {\bf 45}, 2131 (1998); 
E. Paspalakis, C. H. Keitel, and P. L. Knight, ``Fluorescence control through
multiple interference mechanisms", LANL preprint, quant-ph/9810072; 
P. R. Berman, ``An analysis of dynamical supression of spontaneous emission",
LANL preprint, physics/9809005.

\bibitem{nuclear}
Coherence effects in the context of the decay of nuclear levels are discussed
in R. Coussement, M.V. Bergh, G. S'heeren, G. Neyens, and R. Nouwen, 
Phys. Rev. Lett. {\bf 71}, 1824 (1993).

\bibitem{cavity}
For an excellent review and references on theories and experiments on cavity 
induced changes of spontaneous emission, see S. Haroche, and D. Kleppner, 
Phys. Today, January, 1989, p. 24. 
For more recent reviews on this subject, see J.J. Childs, Kyungwon An., 
R.R. Dasari, and M.S. Feld in ``{\it Cavity Quantum 
Electrodynamics}", ed. P.R. Berman (Academic Press, 1993), p. 325.

\bibitem{master}
See the appendix in R.K. Bullough, Hyperfine Interaction {\bf 37}, 71 (1987).

\bibitem{highQ}
E.T. Jaynes, and F.W. Cummings, Proc. IEEE {\bf 51}, 89 (1963);
G.S. Agarwal, J. Opt. Soc. Am. {B\bf 2}, 480 (1985);
H.I. Yoo and J.H. Eberly, Phys. Rep. {\bf 118}, 239 (1985).

\bibitem{spectro} Y.R. Shen, ``{\it The Principles of Non-linear Optics}",
(John Wiley \& Sons, Inc., 1984), chap.13. 

\bibitem{cav-beat}
B. M. Garraway, and P. L. Knight [Phys. Rev. A{\bf 54}, 3592 (1996)] examined
the modifications in quantum beats of a $V$-system, where the system is placed
in a cavity and is {\it prepared} in a coherent superposition of two excited 
states.

\end{thebibliography}
\end{document}